\documentclass[12pt]{article}
\usepackage{graphicx}
\usepackage{cite}

\newcommand\pubnumber{TTK-21-01, P3H-21-001}
\newcommand\pubdate{\today}

\def\institute{
Institute for Theoretical Particle Physics and Cosmology \\  RWTH Aachen
University \\D-52056 Aachen, Germany}
\def\support{\footnote{Work supported in part by the German Research
    Foundation (DFG) Collaborative Research Centre/Transregio project
    CRC/TRR 257: {\it P3H - Particle Physics Phenomenology after the Higgs
    Discovery}.}}

\def\Title#1{\begin{center} {\Large #1 } \end{center}}
\def\Author#1{\begin{center}{ \sc #1} \end{center}}
\def\Address#1{\begin{center}{ \it #1} \end{center}}

\newcommand\pubblock{\rightline{\begin{tabular}{l} \pubnumber\\
         \pubdate  \end{tabular}}}
\newenvironment{Abstract}{\begin{quotation}  }{\end{quotation}}
\newenvironment{Presented}{\begin{quotation} \begin{center} 
             PRESENTED AT\end{center}\bigskip 
      \begin{center}\begin{large}}{\end{large}\end{center} \end{quotation}}

\def\beq{\begin{equation}}
\def\eeq#1{\label{#1}\end{equation}}
\def\eeqn{\end{equation}}

\def\beqa{\begin{eqnarray}}
\def\eeqa#1{\label{#1}\end{eqnarray}}
\def\eeqan{\end{eqnarray}}

\let\bar=\overbar

\def\Dslash{\not{\hbox{\kern-4pt $D$}}}
\def\dslash{\not{\hbox{\kern-2pt $\del$}}}

\def\msb{{\bar{\ssstyle M \kern -1pt S}}}

\begin{document}
\begin{titlepage}
\pubblock

\vfill
\Title{Modelling of top quark decays in $t\bar{t}\gamma$
  production at the LHC}
\vfill
\Author{Malgorzata Worek \support}
\Address{\institute}
\vfill
\begin{Abstract}
  In this proceedings we briefly report on the state-of-the-art NLO
QCD computation for the $pp\to t\bar{t}\gamma$ process in the
di-lepton channel. We describe higher-order corrections to the
$e^+\nu_e \, \mu^- \bar{\nu}_\mu \, b\bar{b}\gamma$ final state at the
LHC with $\sqrt{s}=13$ TeV. Off-shell top quarks,
double-, single- as well as non-resonant top-quark contributions along
with all interference effects are consistently incorporated at the
matrix element level. Results are presented in the form of fiducial
integrated and differential cross sections. The impact of  top
quark modelling is scrutinised by an explicit comparison with the
results in the narrow-width approximation. Both types of
predictions are now  available in the \textsc{Helac-Nlo} framework.

\end{Abstract}
\vfill
\begin{Presented}
$13^\mathrm{th}$ International Workshop on Top Quark Physics\\
Durham, UK (videoconference), 14--18 September, 2020
\end{Presented}
\vfill
\end{titlepage}
\def\thefootnote{\fnsymbol{footnote}}
\setcounter{footnote}{0}
%

%
%
\section{Introduction}
%
%

Top quark pair production in association with an additional photon
provides particularly promising means for a direct measurement of the
top quark electric charge, which governs the coupling strength of the
$t\bar{t}\gamma$ interaction. The $t\bar{t}\gamma$ process can probe,
however, not only the strength but also the structure of the
$t\bar{t}\gamma$ vertex. A generic way for the parameterisation of new
physics effects is provided by an effective field theory (EFT).  The
Lagrangian of the EFT is expressed in terms of the Standard Model (SM)
Lagrangian and a non-SM part, which is dominated by the
contributions due to operators of dimension six. Such contributions
can be translated into top quark magnetic and electric dipole moments
\cite{AguilarSaavedra:2008zc,Schulze:2016qas} and measured at the
LHC. Since the top quark is the heaviest quark, effects of physics
beyond the SM on its couplings are larger than for any other fermion,
thus, deviations with respect to SM predictions for this process
should be easier to detect. Furthermore, $t\bar{t}\gamma$ production
can be used to obtain predictions for integrated and differential
$t\bar{t}\gamma/t\bar{t}$ cross section ratios
\cite{Bevilacqua:2018dny}.  Such cross section ratios are more stable
against radiative corrections and have reduced scale
dependence. Consequently, they have enhanced predictive power. They
can, therefore, be used to study the $t\bar{t}\gamma$ process with a
higher precision. Finally, the integrated and differential top quark charge
asymmetries and lepton charge asymmetry can be investigated in
$t\bar{t}\gamma$ production at the LHC. They provide complementary
information to the measured charge asymmetries in $t\bar{t}$
production \cite{Aguilar-Saavedra:2014vta,
Maltoni:2015ena,Aguilar-Saavedra:2018gfv,Bergner:2018lgm}.

For an accurate comparison with LHC data, reliable theoretical
predictions, which include higher order effects, are mandatory. NLO
QCD corrections for the $t\bar{t}\gamma$ process have been calculated
in Ref. \cite{PengFei:2009ph, Maltoni:2015ena}, whereas NLO
electroweak corrections have been considered in
Ref. \cite{Duan:2016qlc}. In both cases, top quarks are treated as
stable particles. For more realistic studies, however,  top quark decays
are required. First attempts in this direction have been carried out in
Ref. \cite{Kardos:2014zba}, where NLO QCD predictions for
$t\bar{t}\gamma$ have been matched to parton shower programs. In this
study top quark decays are treated in the parton shower approximation
omitting spin correlations and photon emission in parton
shower evolution. Fully realistic theoretical predictions for
$t\bar{t}\gamma$ at NLO in QCD have been presented in Ref.
\cite{Melnikov:2011ta}. In this case, top quark decays are included
using the narrow width approximation (NWA). Consequently, spin
correlations of final state particles are maintained at the NLO level
and photon radiation off top quark decay products is incorporated.
Finally, in Ref. \cite{Bevilacqua:2018woc} a complete description of
$pp\to t\bar{t}\gamma$ in the di-lepton top quark decay channel has
been presented. In this case, all resonant and non-resonant diagrams,
interferences, and off-shell effects of the top quarks and the $W$
gauge bosons are consistently taken into account. These
state-of-the-art theoretical predictions for $t\bar{t}\gamma$  have
been compared to the NWA case in Ref. \cite{Bevilacqua:2019quz} using
the \textsc{Helac-Nlo} common framework \cite{Bevilacqua:2011xh}.

In this proceedings we briefly summarise the state-of-the-art
theoretical predictions for $t\bar{t}\gamma$ production at the LHC and
analyse  the applicability of the NWA approach for this process.

%
%
\section{Results}
%
%

\begin{table}[t!]
  \begin{center}
    \begin{tabular}{|lcc|}
      \hline&&\\
   \textsc{Modelling Approach} & $\sigma^{\rm LO}$ [{\rm fb}]
                              & $\sigma^{\rm NLO}$ [{\rm  fb}]\\
   && \\
  full off-shell & ${7.32}^{+2.45\, (33\%)}_{-1.71\, (23\%)}$
 & ${7.50}^{+0.11\,(1.0\%)}_{-0.45\, (6.0\%)}$  \\[0.2cm] 
        NWA  &
 ${7.18}^{+2.39\,(33\%)}_{-1.68\,(23\%)}$
                              & ${7.33}_{-0.24\,(3.3\%)}^{-0.43\,(5.9\%)}$ \\[0.2cm]
  NWA${}_{\gamma-{\rm prod}}$ 
  & ${3.85}^{+1.29\,(33\%)}_{-0.90\,(23\%)}$  &
    ${4.15}_{-0.21\,(5.1\%)}^{-0.12\,(2.3\%)}$   \\[0.2cm]
  NWA${}_{\gamma-{\rm decay}}$ 
  & ${3.33}^{+1.10\,(33\%)}_{-0.77\,(23\%)}$  &
${3.18}^{-0.31\,(9.7\%)}_{-0.03\,(0.9\%)}$\\[0.2cm] 
        NWA${}_{\rm LOdecay}$  &
                              & ${4.63}^{+0.44\,(9.5\%)}_{-0.52\,(11\%)}$\\[0.2cm]
  \hline
\end{tabular}
\end{center}
\caption{\label{tab:integarted}\it  LO and NLO integrated cross
  sections for various approaches to the modelling of top quark
  decays.  We additionally provide theoretical uncertainties as
  obtained from the scale dependence. All results are provided  for
  $\mu_0=H_T/4$.}
\label{tab:moddeling}
\end{table}
We present results for the $pp \to e^+ \nu_e \, \mu^-
\bar{\nu}_\mu\,b\bar{b}\gamma$ process for the LHC Run II energy of
$\sqrt{s}=13$ TeV. Our calculation uses CT14 parton distribution
functions (PDFs) \cite{Dulat:2015mca} and employs  the following SM
parameters: $G_\mu=1.166378\times 10^{-5} \,{\rm GeV}^{-2}$, $m_W =
80.385$ GeV, $\Gamma_W = 2.0988$ GeV, $m_Z = 91.1876$ GeV and
$\Gamma_Z = 2.50782$ GeV. The electroweak coupling is derived from the
Fermi constant. For the emission of the isolated photon, however,
$\alpha_{\rm QED} = 1/137$ is used instead. The top quark mass is set
to $m_t = 173.2$ GeV. The top quark width is calculated using formulas
from Ref. \cite{Denner:2012yc}. All other QCD partons including $b$
quarks as well as leptons are treated as massless. The final state
jets are constructed using the IR-safe {\it anti-}$k_T$ jet algorithm
\cite{Cacciari:2008gp} with $R = 0.4$. We require at least two jets
for our process, of which exactly two must be bottom flavoured
jets. Furthermore, we request two charged leptons, missing transverse
momentum and an isolated hard photon. For the latter, we use the Frixione
photon isolation prescription, which is based on a modified cone
approach \cite{Frixione:1998jh}. We apply basic selection cuts to
these final states to ensure that they are observed inside the
detector and are well separated from each other, see
\cite{Bevilacqua:2019quz} for more details. We utilise the following
scale $\mu_R = \mu_F=\mu_0 = H_T/4$ where $H_T$ is defined as
\begin{equation}
H_T = p_T(e^+) + p_T(\mu^-) + p_T^{miss}
+p_T(b_1) + p_T(b_2) + p_T(\gamma)\,,
\end{equation}
with $p_T(b_1)$ and $p_T(b_2) $ being bottom-flavoured jets and
$p_T^{miss}$ the missing transverse momentum from the two
neutrinos. The scale systematics is evaluated by varying $\mu_R$ and
$\mu_F$ independently in the range between $\mu_0/2$ and $2
\mu_0$. For comparisons, in the case of the $t\bar{t}\gamma$ cross
section results in the NWA also the fixed scale $\mu_0= m_t/2$ is
used. Our results for the integrated fiducial cross sections for
$t\bar{t}\gamma$ production are summarised in Table
\ref{tab:integarted}.  We compare the full off-shell results with the
calculations in the NWA. For the NWA case two versions are examined:
the full NWA and the NWA${}_{\rm LOdecay}$. The former comprises NLO
QCD corrections to the production and top quark decays as well as
photon radiation from all charged particles. The latter includes the
NWA predictions with LO decays of top quarks and photon radiation in
the production stage only. In Table \ref{tab:integarted} we
additionally quote results for the LO and NLO QCD cross sections where
photon radiation occurs either in the production or in the decay
stage.

A few comments can be made here. First, we observe that the NLO QCD
corrections are small of the order of a few percent only. We can also
assess the size of the non-factorizable top quark corrections for our
setup. These effects, which imply a cross-talk between production and
decays of top quarks, change the NLO cross section by less than $3\%$.
For both the full off-shell and NWA case, theoretical uncertainties
due to the scale dependence are consistently at the $6\%$ when higher
order effects are incorporated. Using results from Table
\ref{tab:integarted}, we can additionally see that $57\%$ of all
isolated photons are emitted in the production stage either from the
initial state light quarks or off-shell top quarks that afterwards go
on-shell. Thus, $43\%$ are emitted in the decay stage, either from
on-shell top quarks or its (charged) decay products. Not only is the
contribution from photon emission in top quark decays substantial but
NLO QCD corrections to decays are also relevant. Therefore, it is not
surprising that NWA${}_{\rm LOdecay}$ results can not reproduce the
correct normalisation. The discrepancy with respect to the full NWA
approach amounts to almost $60\%$. NLO QCD corrections to top quark
decays are negative and at the level of $12\%$ when $\mu_0 = H_T /4$
is employed. Furthermore, theoretical uncertainties increase up to
$11\%$ in this case.
%
\begin{figure}[t!]
  \begin{center}
    \includegraphics[width=0.49\textwidth]{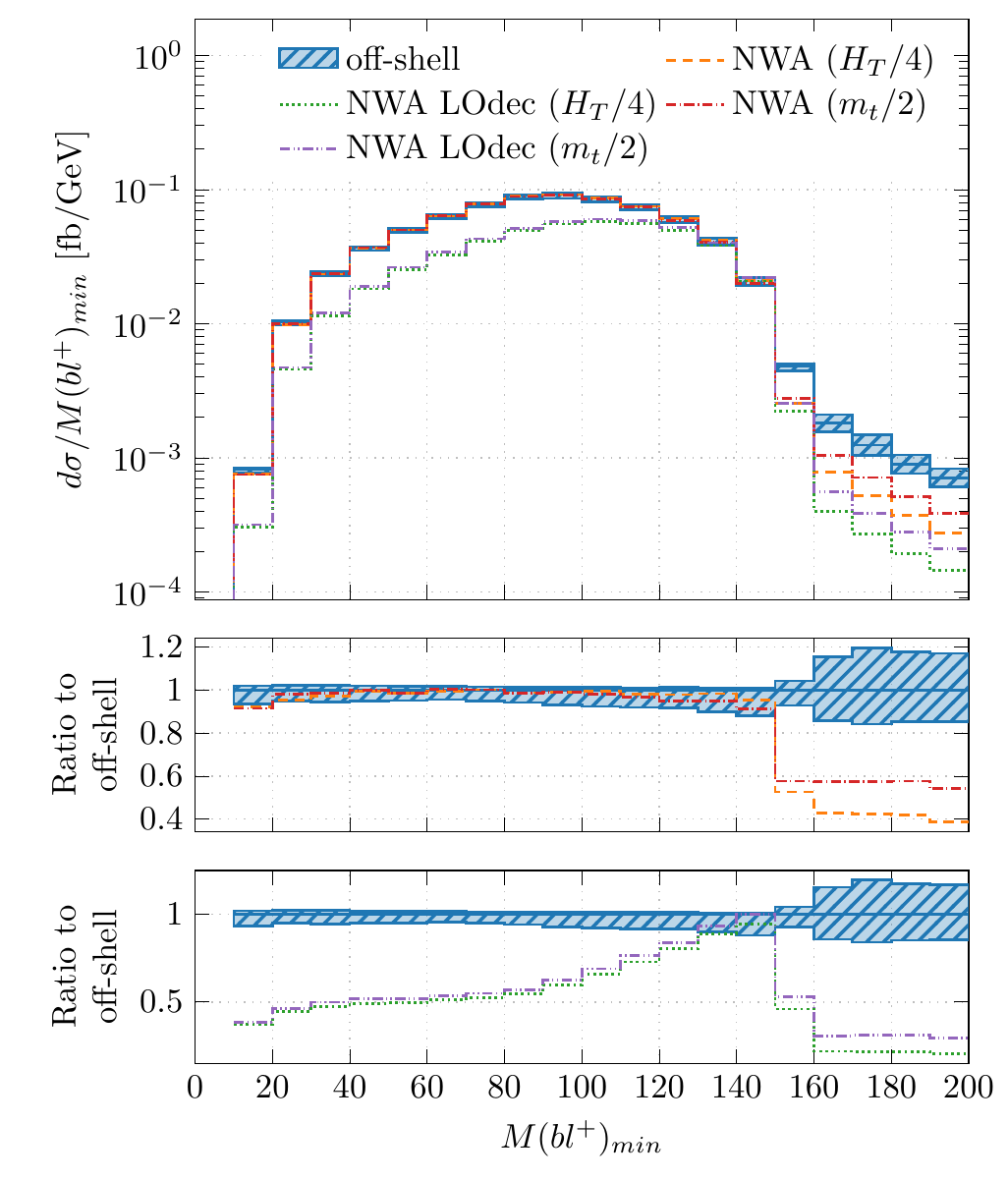}
     \includegraphics[width=0.49\textwidth]{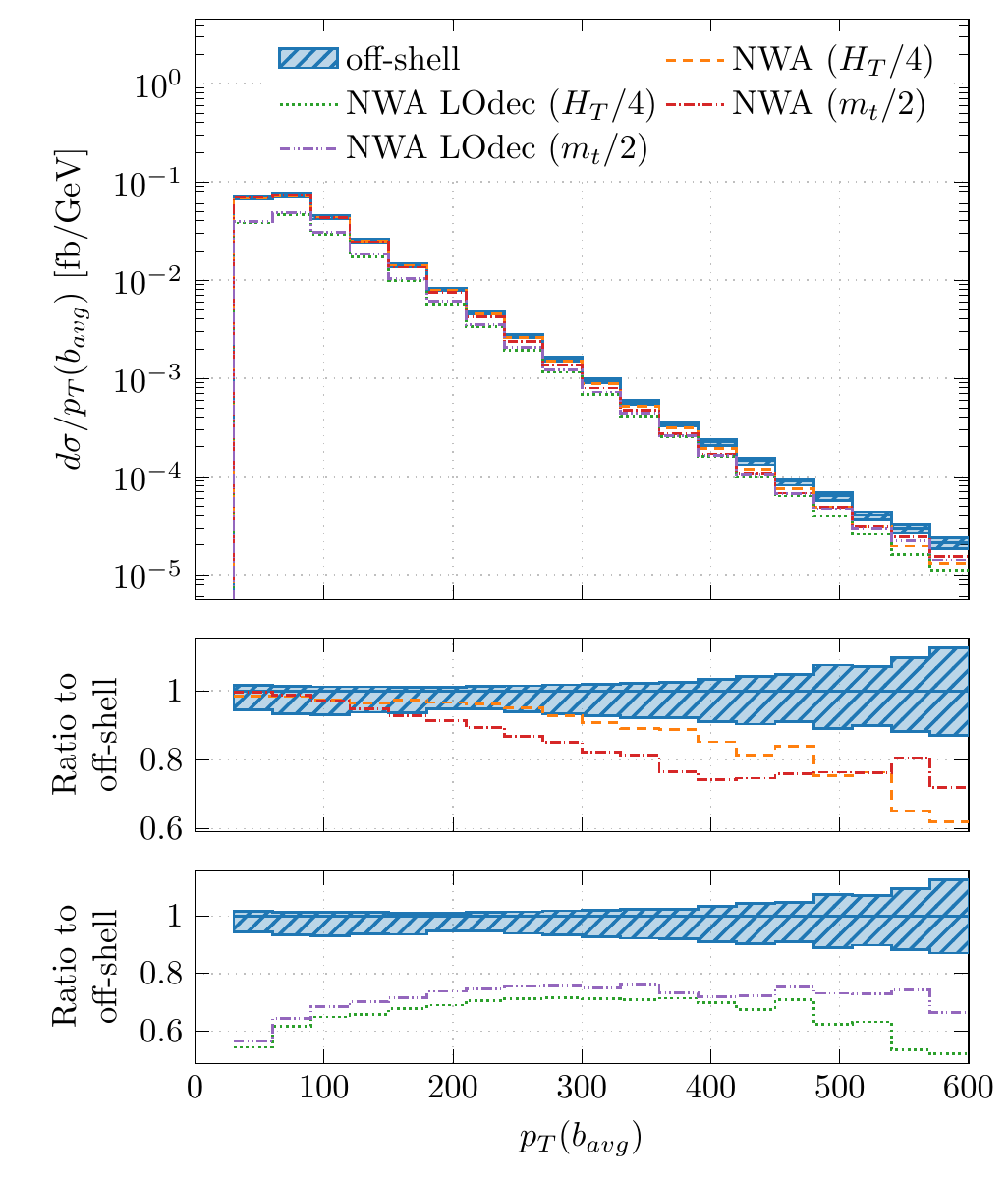}
\end{center}
\caption{\label{fig:1} \it  Differential cross section distributions as
  a function of the minimum invariant mass of the positron and
  bottom-jet and the (averaged) transverse
  momentum of the $b$-jet. NLO QCD
results for various approaches to the modelling of top quark
production and decays are shown.}
\end{figure}
\begin{figure}[t!]
  \begin{center}
    \includegraphics[width=0.49\textwidth]{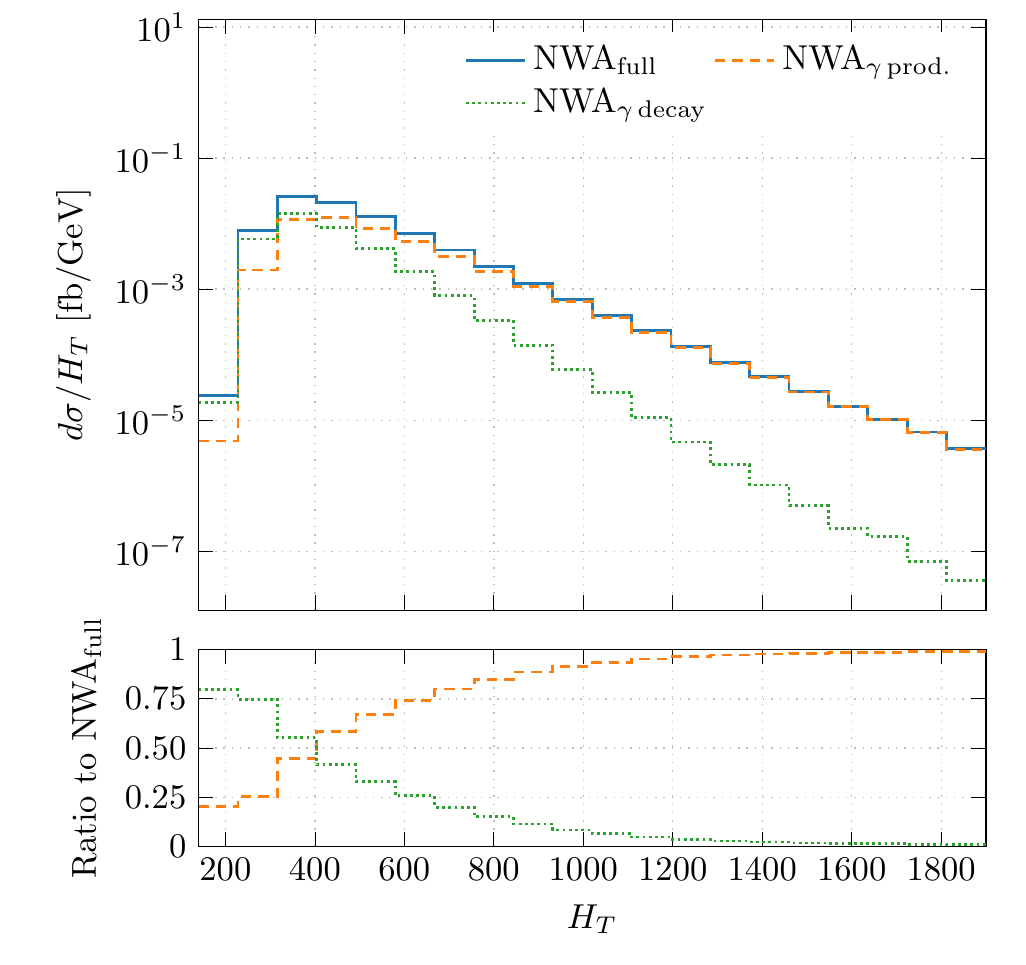}
     \includegraphics[width=0.49\textwidth]{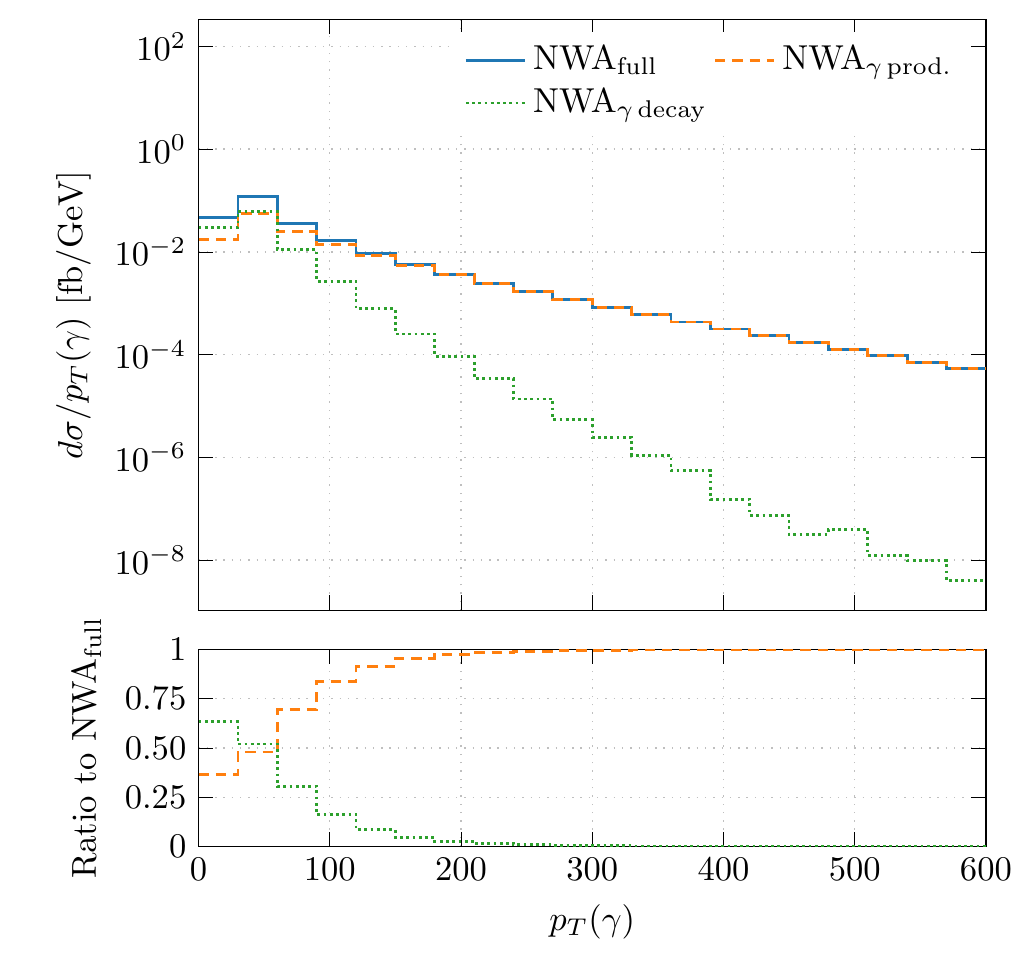}
\end{center}
\caption{\label{fig:2} \it  Differential cross section distributions 
  in function of $H_T$ and the transverse momentum of the isolated
  photon.  NLO QCD predictions in the full NWA are shown together with
  fractions of events originating from photon radiation in the
  production and in decays.}
\end{figure}

The situation changes for the full NWA case once  differential
cross sections are examined instead. In Figure \ref{fig:1}, we present
two examples of differential fiducial cross section distributions
where off-shell effects are particularly sizeable. Specifically,
we show the minimum invariant mass of the positron and bottom-jet and
the (averaged) transverse momentum of the $b$-jet. The upper plots
show absolute NLO QCD predictions for the same three different
theoretical descriptions. The ratios to the full off-shell result
including its scale uncertainty band is plotted in the middle and
bottom plots. When the full off-shell and NWA results are compared,
non-factorizable top quark corrections up to $50\%-60\%$ can be
noticed in the specific phase space regions.  At the same time, we can
see that the NWA${}_{\rm LOdecay}$ predictions are unable to correctly
describe these observables in the whole plotted range. Among all
observables that we have examined, only dimensionful observables are
sensitive to non-factorizable top quark corrections. They can be
divided into two categories: observables with kinematical thresholds
or edges and observables in high $p_T$ regions.

We have also studied the composition of photon emissions at the
differential level. As an example, in Figure \ref{fig:2} differential
cross section distributions in function of $H_T$ and $p_T(\gamma)$
are given at NLO in QCD.  We present predictions in the full NWA
together with fractions of events originating from photon radiation in
the production and in decays. For  low values of the transverse
momentum, the differential distributions are dominated by photon
emission in the decay stage. However, once the high $p_T$ regions of
these observables are probed, photon emission from the production part
of the $t\bar{t}\gamma$ process dominates completely the full
results. Based on our findings, selection criteria can be developed to
reduce such contributions that constitute a background for the
measurement of the anomalous couplings in the $t\bar{t}\gamma$ vertex.

Last but not least, our state-of-the-art theoretical predictions for
the $t\bar{t}\gamma$ process in the $e\mu$ channel at $13$ TeV have 
already been compared to the LHC data   \cite{Aad:2020axn}.

%
%
\section{Summary}
%
%

In this proceedings,  we have briefly described a
comparison between the complete off-shell and NWA calculations for the
$e^+\nu_e \, \mu^-\bar{\nu}_\mu \, b\bar{b}\gamma$ final state at the
LHC. We underlined the importance of the higher order corrections and
photon emission in top quark decays.  Furthermore, we have shown that
dimensionful observables are sensitive to the non-factorizable top
quark corrections in the high $p_T$ regions and close to 
kinematical thresholds or edges. This shows that  proper modelling of
the top quark production and decays is essential already now in the
presence of inclusive cuts and it will become even more important in the
presence of  more exclusive cuts and in the high luminosity phase of
the LHC.

\end{document}